\documentclass[a4paper, amsfonts, amssymb, amsmath, reprint, showkeys, nofootinbib, twoside,superscriptaddress, prb]{revtex4-2}
\usepackage[english]{babel}
\usepackage{here}
\usepackage[utf8]{inputenc}
\usepackage{comment}
\usepackage{bm}
\usepackage{braket}
\usepackage{url}
\usepackage{physics}
\usepackage[colorinlistoftodos, color=green!40,prependcaption]{todonotes}
\usepackage[pdftex, pdftitle={Article}, pdfauthor={Author}]{hyperref}

\newcommand{\m}[1]{\mathcal{#1}}
\newcommand{\mr}[1]{\mathrm{#1}}

\newcommand{\sk}{\sin(k)}

\newcommand{\epk}{\epsilon_k}
\newcommand{\epkp}{\epsilon_{k'}}
\newcommand{\p}{\partial}

\newcommand{\dt}{\frac{\mathrm{d}}{\mathrm{d}t}}

\begin{document}
\title{Asymmetric-Hopping Effective Theory of Current-Carrying Bulk Steady States and Their Effective Temperature}
\author{Yoshihiro Michishita}
\affiliation{Department of Physics, Saitama University, Saitama 338-8570, Japan}
    \date{\today} 
\begin{abstract}
    Current-induced phenomena are often obscured by Joule heating, and their steady states are difficult to analyze in large open systems. We introduce a translationally invariant asymmetric-hopping model as an effective bulk description of boundary-driven systems under current. In a minimal case, it corresponds to an open-system Hatano--Nelson model. In a minimal noninteracting realization, we find that the effective temperature rises linearly with current density in the weak-current regime. The model provides a useful tool for separating intrinsic current-induced effects from heating.
\end{abstract}
\maketitle
\section{Introduction}
Controlling the phase of matter and utilizing its response properties is one of the central missions in condensed matter physics. These efforts have been pursued not only for equilibrium phases of matter but also for nonequilibrium control, such as the control of topological phases of matter\cite{Wang2013-rc, Wintersperger2020-gi}, anomalous Hall effect\cite{McIver2020-dy}, superconductor-like behavior\cite{Fausti2011-wl, Mitrano2016-pc, Liu2020-mx, Budden2021-wo, Buzzi2021-zm, Fava2024-pi}, time crystal\cite{Choi2017-ju, Zhang2017-dl, Liu2025-dg}, and quantum processing by optical control\cite{Tomita2017-oo, Noel2022-uq, Google-Quantum-AI-and-Collaborators2023-cr}. 
One important direction is the electric-current control of material properties, which can induce metal-insulator transition
\cite{Zhang2019-qb, Suen2024-id}, magnetic phase transition or manipulation\cite{Jain2006-om, Wadley2016-fj, Grzybowski2017-zi, Hrabec2017-as, Bodnar2018-sx, Arpaci2021-si, Reimers2023-kv, Zheng2024-in, Shimizu2025-fx}, metastable states\cite{Huber2025-ro}, and pattern formation\cite{Gauquelin2023-in}. 
On the other hand, electric-current control often faces the experimental difficulties caused by Joule heating. Recently, some experiments reported on current-induced phenomena have been  subject to renewed scrutiny\cite{Kitaori2021-jc, Furuta2024-hy, Cao2025-uj, Zhang2025-vc}, which is mainly caused by the heating of background equipment and the sample itself. To avoid this problem, several studies have attempted to suppress or analyze heating\cite{Mattoni2020-bh, Zhang2019-qb, Curcio2023-rl, Mattoni2024-sh, Ootsuki2025-em, Fang2025-lh, Liu2026-us}, for example by adding the sample to a heat bath with a large heat capacity. However, Ref.\cite{Ootsuki2025-em} pointed out that the estimated effective temperature of the sample becomes higher than the bath temperature, and the increase of the effective temperature seems linearly proportional to the current density\footnote{see figure.2(b) in Ref.\cite{Ootsuki2025-em}}, while, within the simple analysis, the Joule heating cannot reproduce linear dependence\cite{Supple}. Analyzing the steady state under current and its effective temperature is essential to distinguish the intrinsic current-induced phenomena from the heating effect or the increase of effective temperature.

In theoretical treatments of isolated systems driven by an electric field, within either linear or nonlinear response theory, temperature is generally introduced as an external parameter. Consequently, these approaches do not offer a direct procedure for evaluating the effective temperature of a system carrying a current. Determination of the effective temperature instead requires coupling the driven system to a heat bath and formulating it as an open system, in which the heat generated by the current is compensated by heat flow into the bath under steady-state conditions.

Recently, much effort has been made to analyze the steady state in the boundary-driven system (BDS)\cite{Wichterich2007-mg, Prosen2008-bm, Prosen2011-wp, Prosen2014-xx, Landi2022-gk, Kempa2026-fm} and it is revealed that the effective temperature of the expectation values of physical quantities is sometimes different from the bath temperature\cite{Landi2022-gk}. Although some models can be solved exactly, the analysis of the steady state of the BDS is generally difficult, and the numerical analysis is often used. However, the numerical analysis of the BDS is also difficult because of the large Hilbert space of the system and the bath. Thus, effective models with translational symmetry would be much simpler and useful to analyze the effective bulk steady state under current.

In this paper, we introduce a translationally invariant asymmetric-hopping model as an effective description of the bulk steady state of a boundary-driven system under current. For the noninteracting fermionic model considered here, we analyze its steady state and effective temperature and show that, in the weak-current regime, the increase in the effective temperature is linearly proportional to the current density, which shows behavior similar to the experimental observation in Ref.~\cite{Ootsuki2025-em}.

\section{Brief Introduction to Boundary-Driven System}
\begin{figure}[t]
\includegraphics[width=0.45\linewidth]{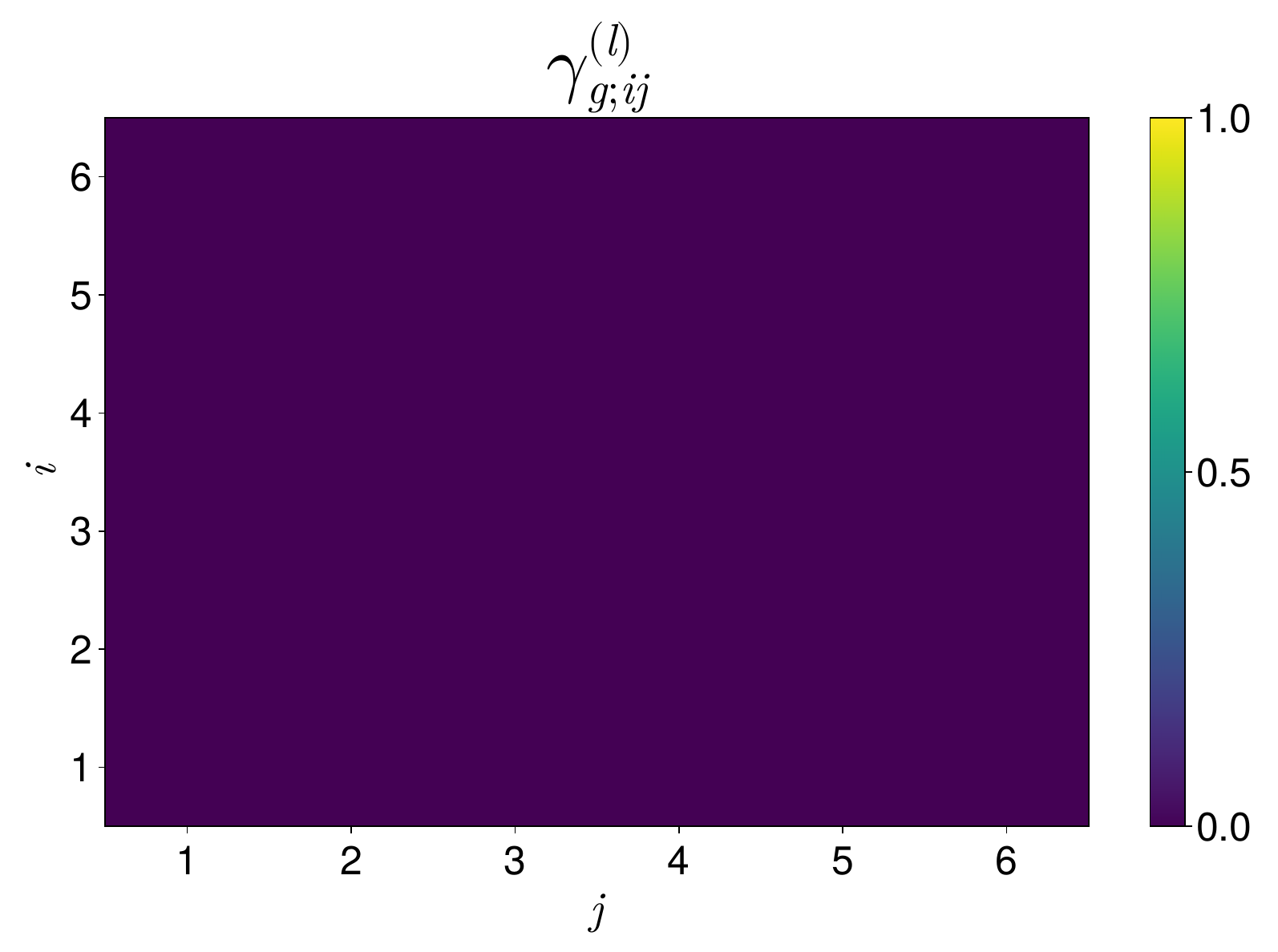}
\includegraphics[width=0.45\linewidth]{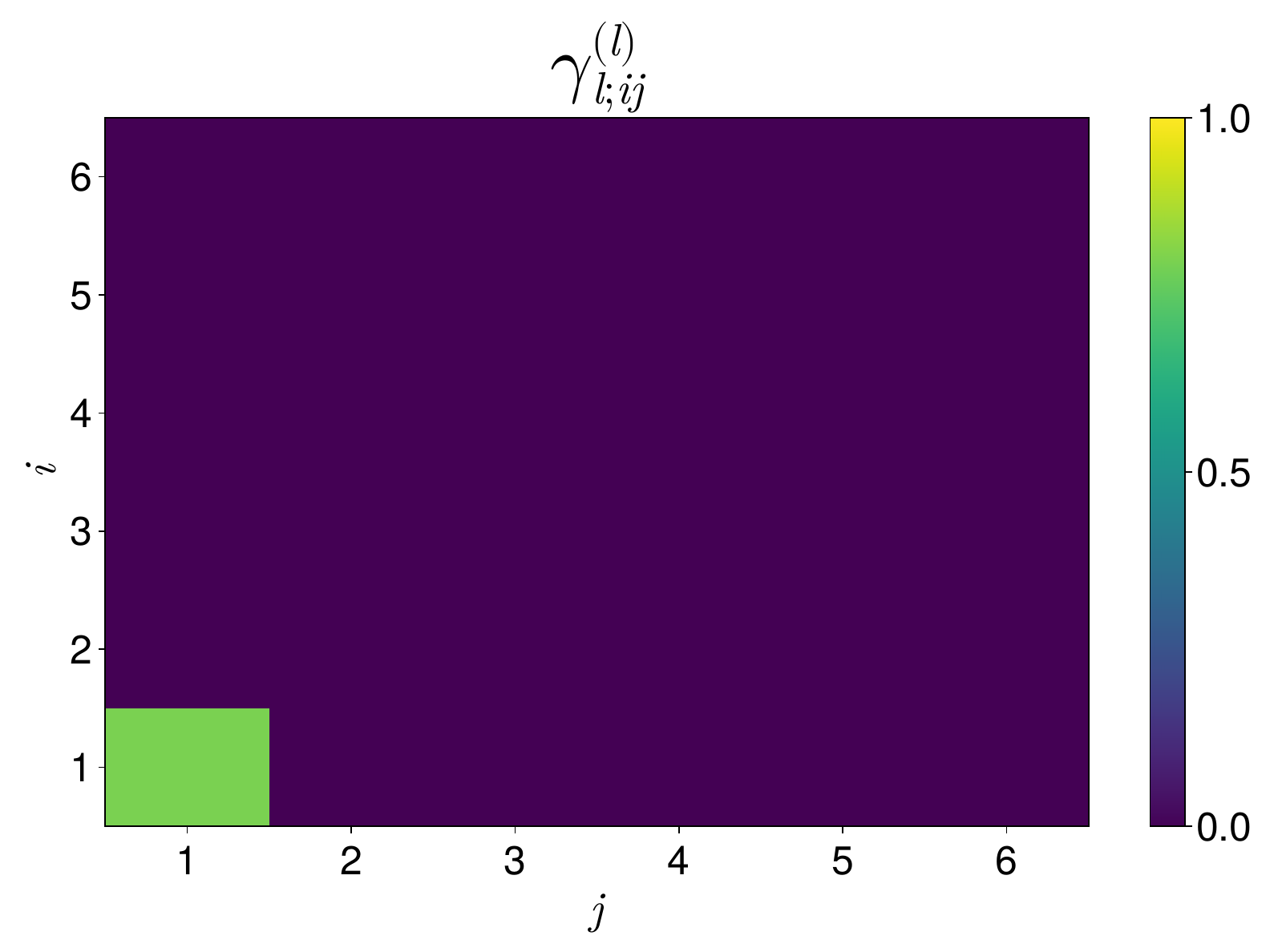}
\includegraphics[width=0.45\linewidth]{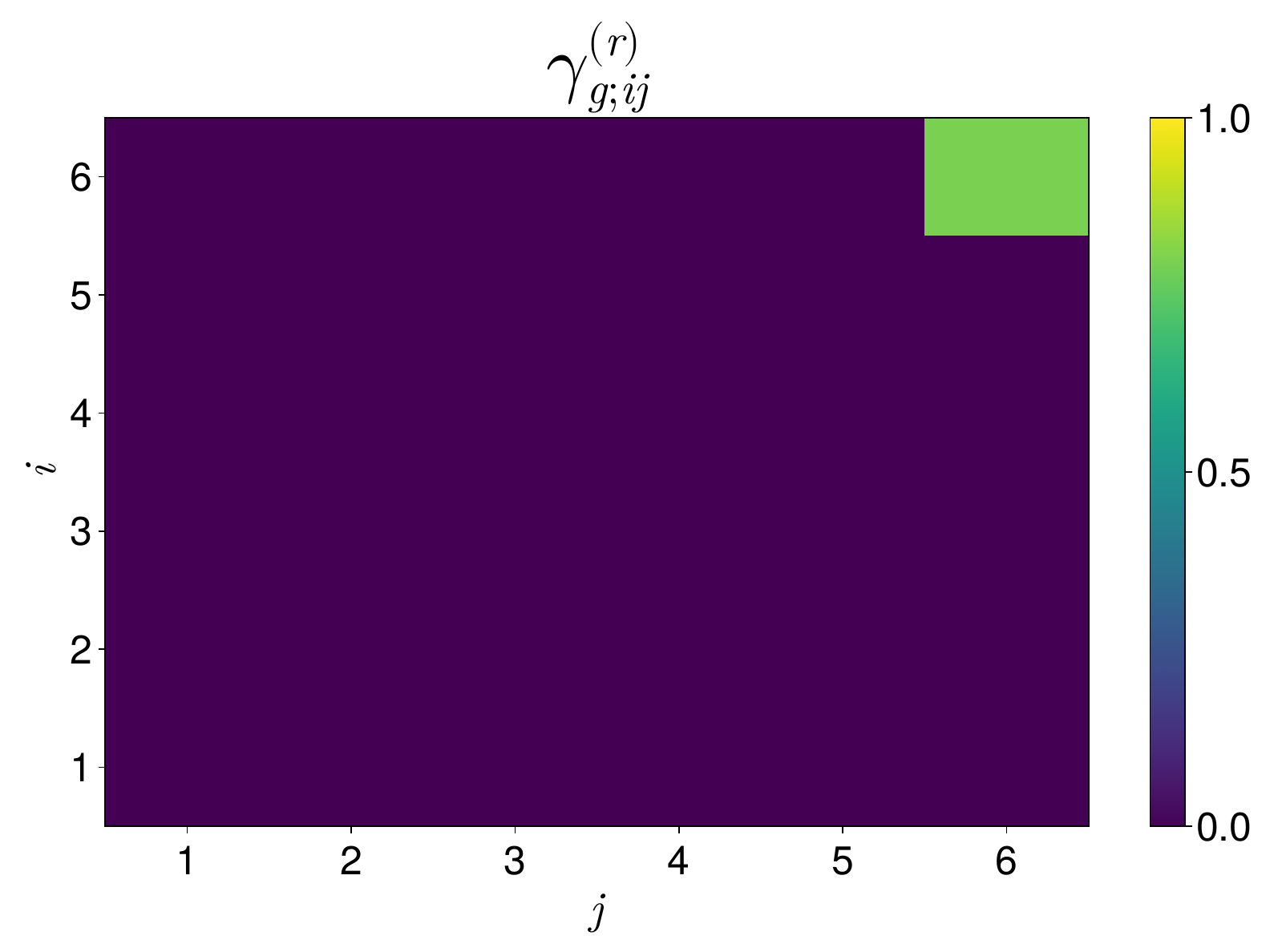}
\includegraphics[width=0.45\linewidth]{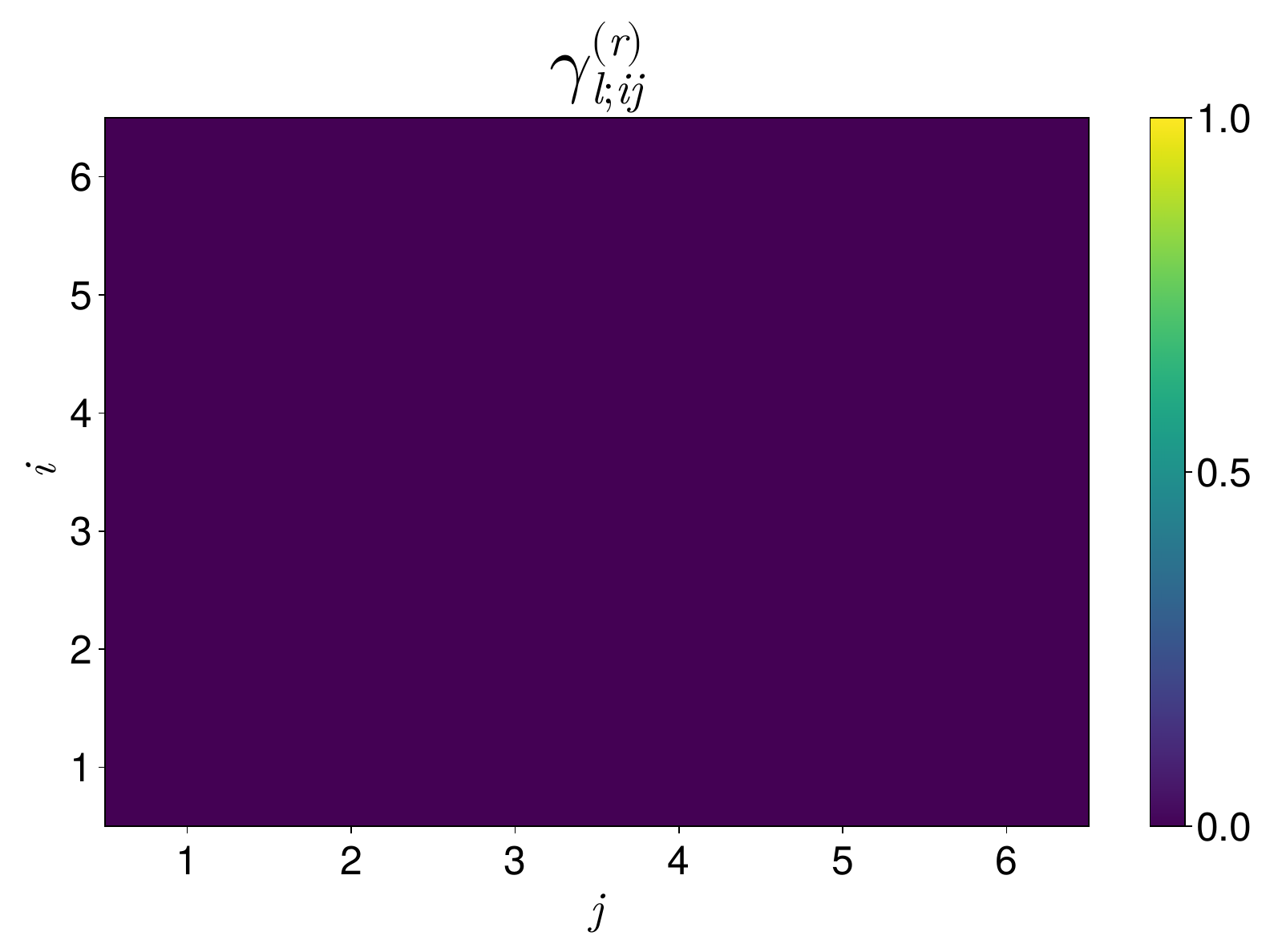}
\caption{The gain and loss terms in the Lindblad equation for the boundary-driven system for $l=6$. The heatmaps show the coefficients of the jump terms for gain ($c^\dagger_i \rho c_j$) and loss ($c_i \rho c^\dagger_j$). See the supplementary material for detailed bath parameters.
\label{fig:Fig1}}
\end{figure}

\begin{figure}[h]
\includegraphics[width=0.46\linewidth]{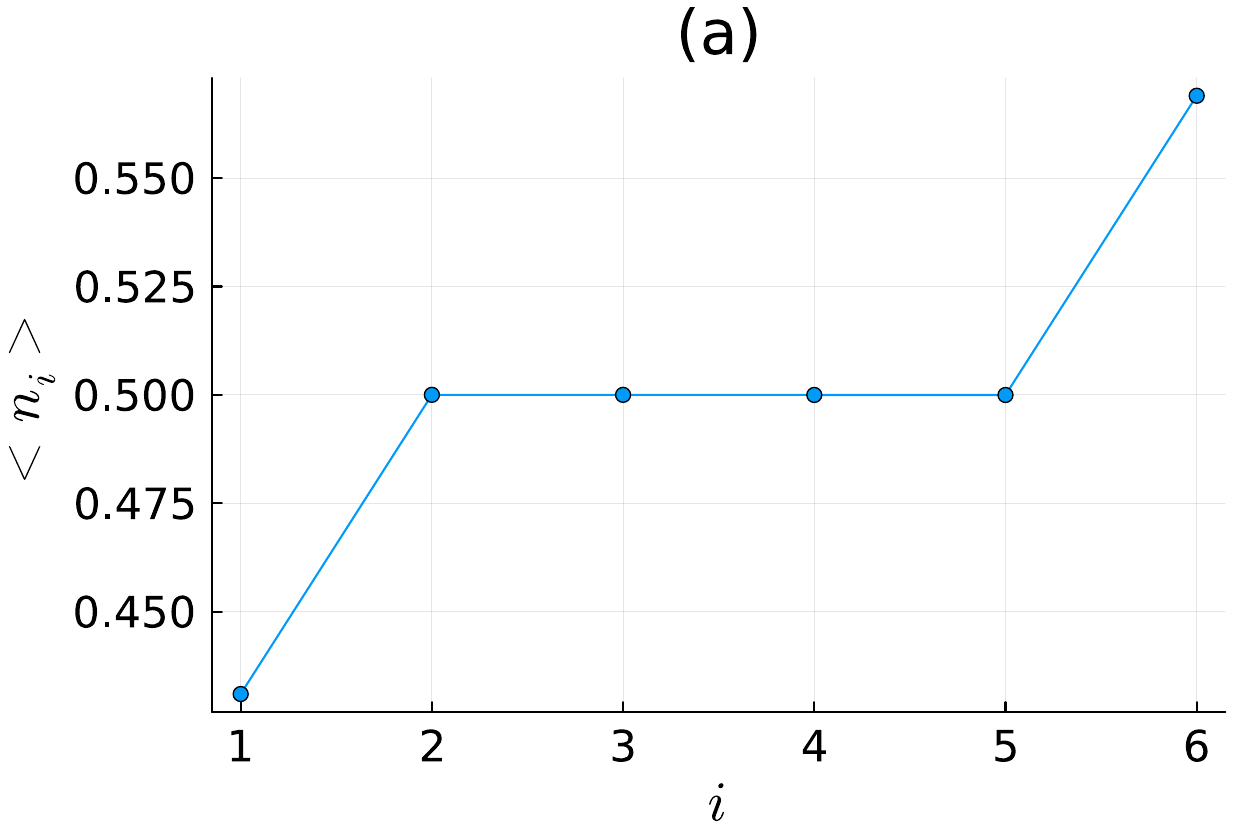}
\includegraphics[width=0.46\linewidth]{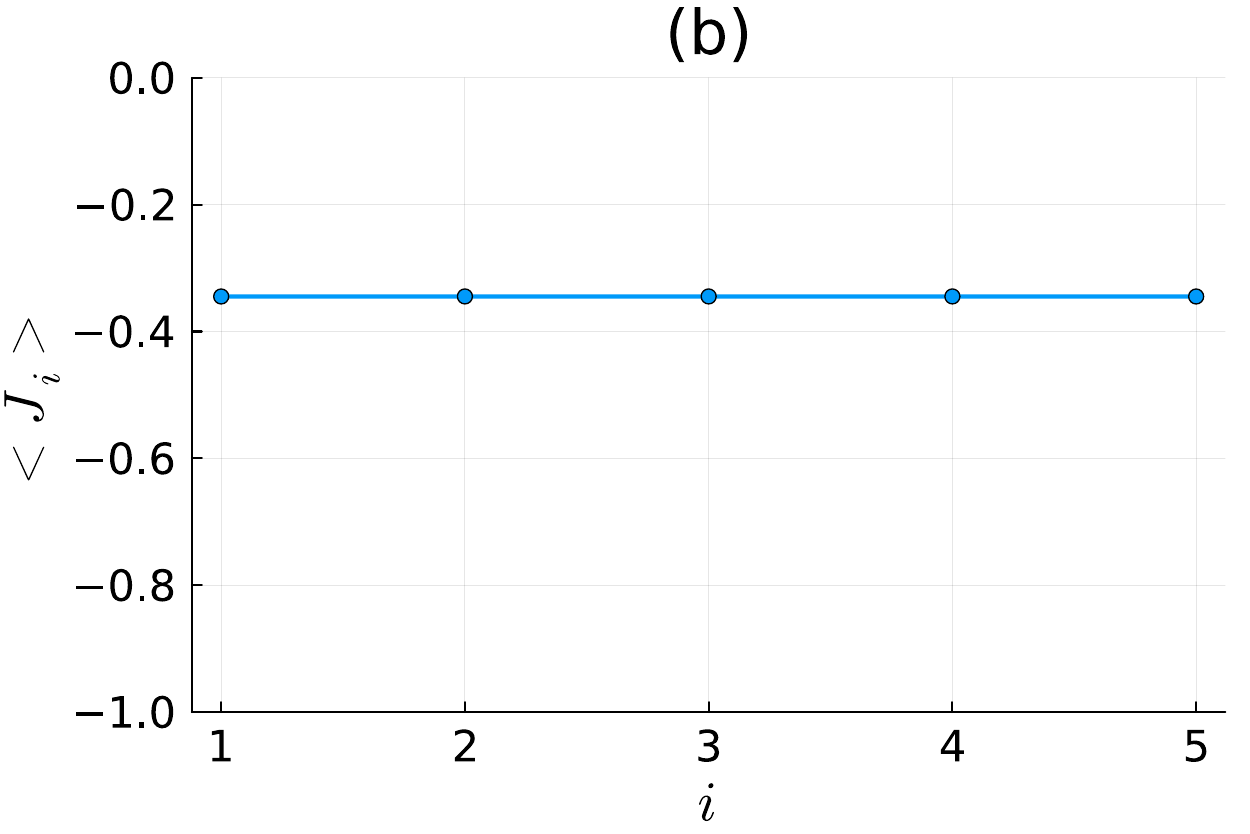}
\includegraphics[width=0.46\linewidth]{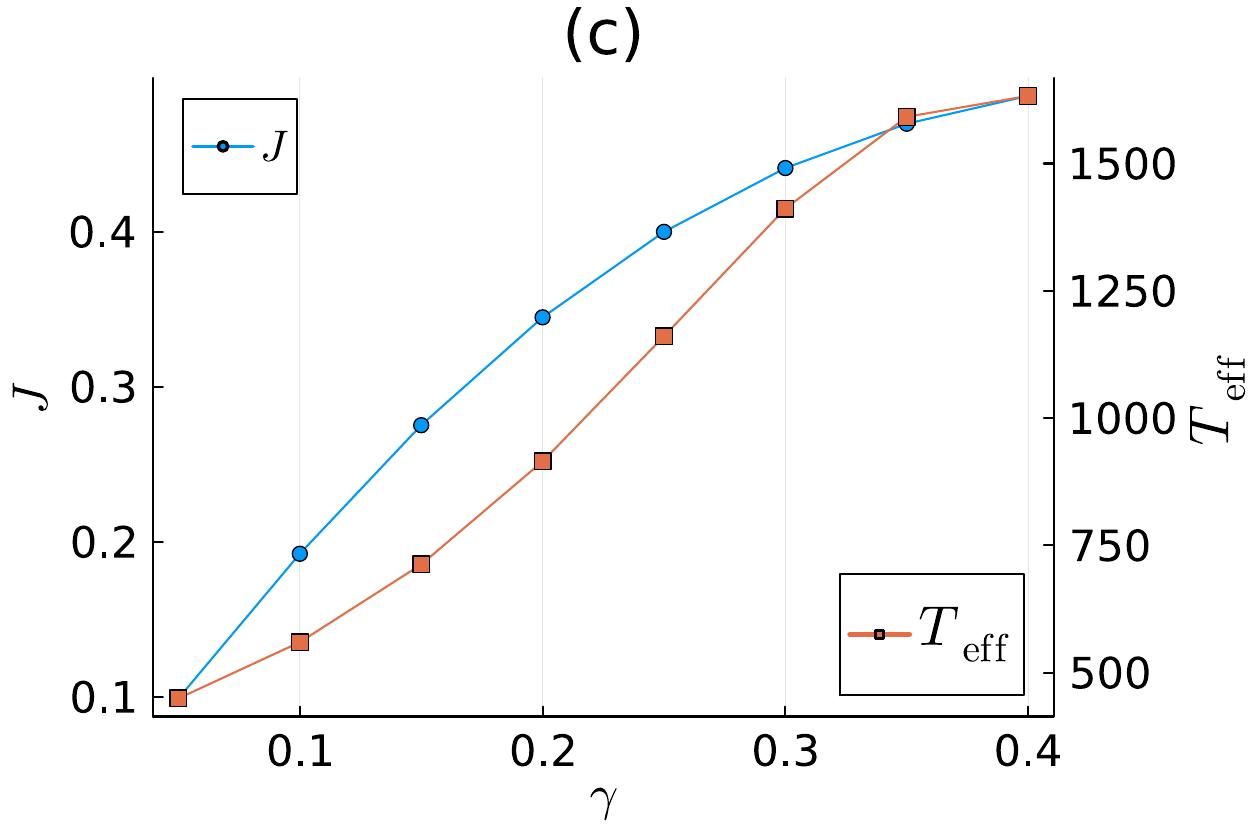}
\includegraphics[width=0.46\linewidth]{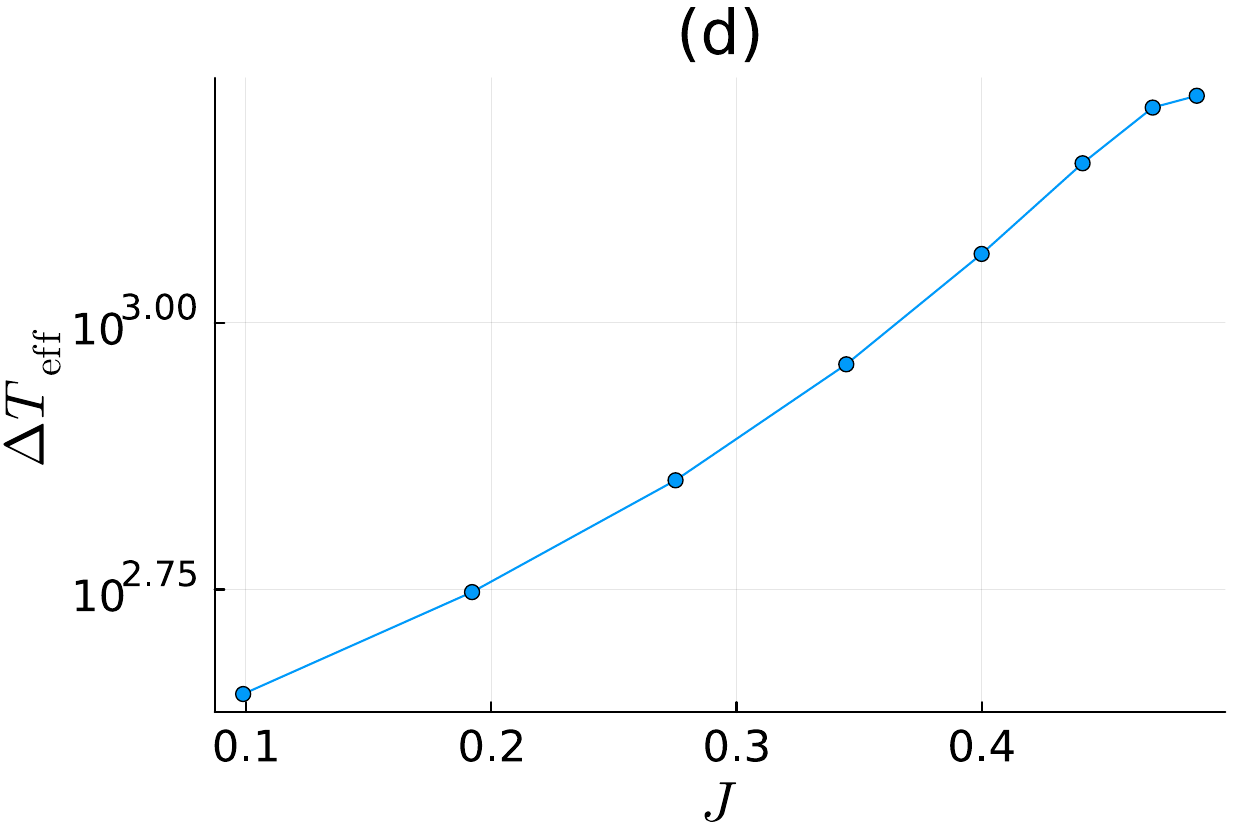}
\caption{Physical quantities in the bulk region in the steady state of the BDS. The upper panel shows the site-dependence of (a) the local occupation and (b) the local current in the steady state. Both quantities are uniform in the bulk region. The lower panel shows (c) the current and the effective temperature as a function of $\gamma$, which is the strength of the coupling between the system and the edge current source. The current and the effective temperature at the central two sites in the steady state.  Panel (d) shows the relation between the current and effective temperature on a semilogarithmic scale. The effective temperature increase exhibits an approximately exponential-like dependence at larger currents but becomes approximately linear in the small current regime.
\label{fig:Fig2}}
\end{figure}

We first briefly analyze the particle distribution, the effective temperature, dissipation, and current in the bulk region in BDS as an introduction to the next section. Hereafter, we set the reduced Planck constant $\hbar$, the lattice constant $a$, Boltzmann constant $k_\mathrm{B}$, and the electron charge $e$ to unity. In this section, we examine the following properties: (i) the jump terms only appear at the edges, (ii) the particle distribution is uniform in the bulk, and (iii) the local distribution has higher effective temperature than the edge bath, and these properties are essential for deriving the effective model for the bulk region. 
For simplicity, we consider a spinless free-fermion model coupled to two different baths.
Rather than introducing the dissipative coefficients phenomenologically, we derive them from the system--bath coupling within the universal Lindblad equation (ULE)\cite{Nathan2020-mj,Davidovic2022-zp,Nathan2024-ag,Ikeuchi2025-wd}, thereby allowing us to examine whether the above properties emerge in a physically motivated boundary-driven setup.
We use the ULE because it guarantees the complete positivity without the secular approximation, which can affect the bulk charge transport\cite{Wichterich2007-mg, Landi2022-gk}. The resulting Lindblad equation reads
\begin{align}
    \dt \rho(t) &= -i[\m{H},\rho(t)] + \m{D}^{(l)}[\rho] + \m{D}^{(r)}[\rho].\label{NR}\\
    \m{H} =& -t\sum_{j=1}^{l-1} (c_j^{\dagger} c_{j+1} + c_{j+1}^{\dagger} c_j),\label{Ham_boundary}
\end{align}
\begin{align}
    \m{D}^{(r/l)}[\rho] &= \sum_{i,j} \Bigl[\gamma^{(r/l)}_{l;ij}\Bigl(c_i \rho c_j^{\dagger} - \frac{1}{2}\{c_j^{\dagger}c_i, \rho\}\Bigr)\nonumber\\
    & \ \ \ \ \ \ +\gamma^{(r/l)}_{g;ij}\Bigl(c_i^{\dagger} \rho c_j - \frac{1}{2}\{c_jc_i^{\dagger}, \rho\}\Bigr)\Bigr],\label{dissipator_boundary}
\end{align}
where $c^{(\dagger)}_{i}$ is the annihilation (creation) operator for the spinless fermion at site $i$, $\m{H}$ is the system Hamiltonian, $t$ is the hopping parameter, $\m{D}^{(l/r)}$ are the dissipators caused by the coupling to the left and right thermal baths, $\gamma^{(r/l)}_{g;ij}$ and $\gamma^{(r/l)}_{l;ij}$ are the coefficients of the gain and loss terms, respectively, obtained from the coupling to the right/left fermionic bath through the ULE.
The matrices $\gamma^{(r/l)}_{g}$ and $\gamma^{(r/l)}_{l}$ are positive semidefinite, as required for complete positivity of the Lindblad dynamics.
Fig.~\ref{fig:Fig1} displays $\gamma^{(r/l)}_{g;ij}$ and $\gamma^{(r/l)}_{l;ij}$ in Eq.~\ref{NR}, calculated for $T^{(\mr{l/r})}_{\mr{B}}=0.05$ as a representative low-temperature value compared with the hopping scale $t=1$.
We set the chemical potentials of the left and right baths below the band bottom and above the band top, respectively. Details of the bath setup and the explicit construction of these coefficients are provided in the Supplemental Material.
Because the jump terms are strongly localized near the edges, charge conservation ensures that the local charge current remains constant throughout the bulk. We note that jump terms in the Lindblad representation can generally be nonlocal even when the system--bath coupling is localized at the edge. Thus, the localization observed here is nontrivial and results from the present universal Lindblad equation setup.
Fig.~\ref{fig:Fig2}(a) and (b) show the site-dependence of the local occupation and the local current in the steady state of the BDS. 
In the bulk region, where the jump terms are negligible, the local particle density $n_i=c_i^\dagger c_i$ satisfies the continuity equation $\mr{d}\ev{n_i}/\mr{d}t=\ev{J_{i-1}}-\ev{J_i}$, where $J_i=-it(c_{i+1}^\dagger c_i-c_i^\dagger c_{i+1})$ is the local current operator on the bond between sites $i$ and $i+1$. Thus, the steady-state condition $d\langle n_i\rangle/dt=0$ requires $\langle J_{i-1}\rangle=\langle J_i\rangle$, explaining the spatially uniform current in the bulk. The local occupation is also approximately uniform in the bulk region, as shown in Fig.~\ref{fig:Fig2}(a).

Fig.~\ref{fig:Fig2}(c) and (d) show the current and the local two-site effective temperature at the central two sites in the steady state as a function of $\gamma$, which is the strength of the coupling between the system and the edge current source. We note that both $\gamma^{(r/l)}_{g;ij}$ and $\gamma^{(r/l)}_{l;ij}$ are proportional to $\gamma$ in the Nathan--Rudner construction. 
In a nonequilibrium state, the local effective temperature is generally not uniquely defined. For example, it can be determined either by fitting the local distribution to a thermal distribution or by matching the local energy density to that of a thermal state. Here, we adopt the former definition and determine the effective temperature by fitting the occupation distribution of the central two-site subsystem to a Fermi distribution. In the present case, the two definitions give the same result, as discussed in the Supplemental Material~\cite{Supple}.
We note that the limit $\gamma\rightarrow0$ in the present boundary-driven setup should not be identified with an equilibrium limit. The chemical-potential difference between the left and right baths is kept fixed while the overall system--bath coupling $\gamma$ is reduced. Consequently, the current-driving and thermalizing effects of the reservoirs vanish simultaneously, and the steady state is not guaranteed to approach a thermal equilibrium state at the bath temperature. Consistent with this, Fig.~\ref{fig:Fig2}(c) shows that the locally defined effective temperature does not approach $T_{\rm B}=0.05$ as $\gamma\rightarrow0$.
The current and the effective temperature increase as $\gamma$ increases, and the effective temperature exhibits an approximately exponential dependence on the current and is higher than the bath temperature. Although the effective temperature shows a nonlinear dependence on the current over a wide range of the driving strength, its small-current behavior is approximately linear. This weak-driving regime can be analyzed perturbatively with an effective model as we show in the following sections.
 
In constructing an effective Liouvillian for the bulk regime, we require that it reproduce the following local properties of the boundary-driven steady state: (i) a translationally uniform density in the bulk and (ii) a finite, spatially uniform local current. These local signatures do not uniquely determine the effective Liouvillian. Nevertheless, as shown below, they impose a strong constraint on the form of the local jump operators: a dissipative term containing an asymmetric hopping component is required to sustain a local steady state with finite current at uniform density. In addition, we have shown that the locally defined effective temperature in the bulk is much higher than the bath temperature. This observation motivates the simplest representative form of the jump operators introduced below.

\section{Construction of the Effective Model \label{sec:model}}
Here we introduce an effective model for the bulk region of the BDS whose steady state has finite local current and uniform particle distribution. For simplicity, we consider below the same model as in the previous section, although the following discussion is not restricted to one dimension.
We start from the steady state of the BDS which we have analyzed in the previous section as
\begin{align}
\dt \rho_{\mr{ss}} =& \m{L}\rho_{\mr{ss}} = -i[\m{H},\rho_{\mr{ss}}] + \m{D}_{\mr{BD}}[\rho_{\mr{ss}}] = 0,
\end{align} 
where $\m{D}_{\mr{BD}}$ is the dissipator for the BDS and it corresponds to the sum of the boundary localized dissipators $\m{D}^{(r)} + \m{D}^{(l)}$ in Eq.~(\ref{dissipator_boundary}). We can define the bulk steady state and construct the effective Liouvillian $\m{L}_{\mr{eff}}$ which has the same bulk steady state as
\begin{align}
    \rho^{(\mr{bulk})}_{\mr{ss}} &= \mr{Tr_{edge}}\Bigl[ \rho_{\mr{ss}} \Bigr],\\
    \m{L}_{\mr{eff}} \rho^{(\mr{bulk})}_{\mr{ss}} &= -i[\m{H}^{(\mr{bulk})}, \rho^{(\mr{bulk})}_{\mr{ss}}] + \m{D}_\mr{eff}[\rho^{(\mr{bulk})}_{\mr{ss}}] = 0,\label{effective_Liouvillian}
\end{align}
where \(H^{(\mathrm{bulk})}\) denotes the local Hamiltonian terms supported inside the bulk region.
We note that, in general, we cannot uniquely determine $\m{D}_\mr{eff}$($\m{L}_{\mr{eff}}$) only from Eq.~(\ref{effective_Liouvillian}) and additional restrictions are required to determine $\m{L}_{\mr{eff}}$.

While direct analysis of the BDS becomes difficult for large systems, a translationally invariant effective description is useful when the bulk steady state exhibits spatially uniform local observables, since it captures the bulk properties without explicitly retaining the boundary structure.
For this purpose, we assume that the effective dissipative dynamics can be represented by local jump operators. Because the local particle density is a one-site quantity whereas the local current is defined on a bond, a two-site subsystem is the minimal local unit that can characterize both properties. We therefore consider a two-site effective dissipator $\m{D}^{(j)}_{\mr{eff}}$ acting on neighboring sites $j$ and $j+1$, and require it to reproduce the reduced two-site steady state of the BDS through the local steady-state condition in Eq.~(\ref{effective_Liouvillian3}). 
We then construct the effective dissipator for the whole bulk region by applying the same local dissipative structure to every bond and summing over $j$. This construction yields a translationally invariant bulk model while preserving the local steady-state properties extracted from the original BDS. The translational invariance also makes the resulting bulk model analytically tractable in momentum space, allowing us to study the current-carrying steady state without explicitly treating the boundaries. As shown below, the simplest realization of this construction is directly connected to the Hatano--Nelson model, providing a simple interpretation of the asymmetric-hopping structure.

The local condition reads
\begin{align}
    -i[-t(c_j^{\dagger} c_{j+1} + c_{j+1}^{\dagger} c_j), \rho^{j, j+1}_{\mr{ss}}] + \m{D}^{(j)}_{\mr{eff}}[\rho^{j, j+1}_{\mr{ss}}] = 0,\label{effective_Liouvillian3}
\end{align}
where $\rho^{j, j+1}_{\mr{ss}} = \mr{Tr_{rest}}\Bigl[ \rho^{(\mr{bulk})}_{\mr{ss}} \Bigr]$ is the local two-site steady state for the sites $j$ and $j+1$, which can be obtained by tracing out the rest of the bulk system. 
We then construct a translationally invariant effective Liouvillian for the whole bulk region as
\begin{align}
    \tilde{\m{D}}_{\mr{eff}}[\rho] &= \sum_j \sum_{\alpha}\m{D}^{(j, \alpha)}_{\mr{eff}}[\rho]\nonumber\\
    \m{D}^{(j, \alpha)}_{\mr{eff}}[\rho] &= \gamma^{(l)}_{\alpha} \Bigl(L^{(l)}_{(j,\alpha)}\rho L^{(l)\dagger}_{(j,\alpha)} - \frac{1}{2}\{L^{(l)\dagger}_{(j,\alpha)} L^{(l)}_{(j,\alpha)}, \rho\}\Bigr)\nonumber\\
    & \quad + \gamma^{(g)}_{\alpha} \Bigl(L^{(g)}_{(j,\alpha)}\rho L^{(g)\dagger}_{(j,\alpha)} - \frac{1}{2}\{L^{(g)\dagger}_{(j,\alpha)} L^{(g)}_{(j,\alpha)}, \rho\}\Bigr),\label{effective_Liouvillian2}\\
    L^{(l)}_{(j,\alpha)} &= \alpha_1 c_j + \alpha_2 c_{j+1},\quad L^{(g)}_{(j,\alpha)} = \alpha_3 c_j^{\dagger} + \alpha_4 c_{j+1}^{\dagger},\label{jump_gain}
\end{align}
where $\m{D}^{(j)}_{\mr{eff}} = \sum_\alpha\m{D}^{(j,\alpha)}_{\mr{eff}}$. We note that $\tilde{\m{D}}_{\mr{eff}}$, in general, does not satisfy the steady state condition in Eq.~(\ref{effective_Liouvillian}) for the whole bulk region, but it is constructed to satisfy the local consistency condition in Eq.~(\ref{effective_Liouvillian3}) and preserves the translational symmetry.

Interestingly, when the local steady state has a uniform site-resolved particle density and a finite local current,
\begin{align}
\ev{\m{J}} = \mr{Tr}_{j, j+1}[-it(c^\dagger_{j+1}c_j-c^\dagger_j c_{j+1})\rho^{j, j+1}_{\mr{ss}}]\neq 0,\label{local_current}
\end{align}
as we have shown in the previous section, the condition in Eq.~(\ref{effective_Liouvillian3}) requires a finite imaginary part of $\alpha_1\alpha_2^*$ and/or $\alpha_3\alpha_4^*$. (see the Supplemental Material for a detailed derivation.\cite{Supple})
This term can be interpreted as an asymmetric hopping term because the effective non-Hermitian Hamiltonian of this model can be written as
\begin{align}
    &\m{H}^{(j,j+1)}_{\mr{eff}}\nonumber\\
    &= \m{H}^{(j,j+1)} -\frac{i}{2} \sum_{\alpha} \Bigl[\gamma^{(l)}_{\alpha} L^{(l)\dagger}_{(j,\alpha)} L^{(l)}_{(j,\alpha)} + \gamma^{(g)}_{\alpha} L^{(g)\dagger}_{(j,\alpha)} L^{(g)}_{(j,\alpha)}\Bigr]\nonumber
\end{align}
and the imaginary part of $\alpha_1\alpha_2^*$ and/or $\alpha_3\alpha_4^*$ appears as the coefficient of the asymmetric hopping term $(c^\dagger_j c_{j+1} - \mr{h.c.})$ in the effective non-Hermitian Hamiltonian.
This construction of the  effective model should be understood as a minimal local representation of the bulk steady state rather than a unique derivation of the full reduced Liouvillian. 
We use the uniform density and finite current obtained in the bulk of the BDS as constraints to determine the form of the effective model, resulting in the asymmetric hopping term. We then choose the simplest representative jump operators satisfying this condition and analyze their steady state after adding a thermal relaxation channel.
 
Below, we consider a single $\alpha$ channel and set $(\alpha_1, \alpha_2, \alpha_3, \alpha_4) = (1, -i, 1, i)$ as the simplest choice to satisfy the above conditions for the finite local current. We note that we can partially justify this choice of $\alpha$ when the effective local temperature is much higher than the hopping amplitude.\cite{Supple} 

Under the periodic boundary condition where we set $c_{l+1} = c_1$, we can summarize our model in wave number space as

\begin{align}
&\tilde{\m{D}}_{\mr{eff}}[\rho(t)] \nonumber\\
&=\sum_k \Biggl[\gamma^{(l)}(\epsilon_k) (2-2\sin(k))\Bigl(c_k\rho(t)c_k^{\dagger} - \frac{1}{2}\{c_k^{\dagger}c_k, \rho(t)\}\Bigr)\nonumber\\
&\quad\quad +\gamma^{(g)}(\epsilon_k) (2+2\sin(k))\Bigl(c_k^{\dagger}\rho(t)c_k - \frac{1}{2}\{c_kc_k^{\dagger}, \rho(t)\}\Bigr)\Biggr].\label{jump_current2}
\end{align}
We note that, for $\gamma^{(l)}(\epsilon_k)=\gamma^{(g)}(\epsilon_k)=h$, the effective Hamiltonian of this model becomes equivalent to the Hatano--Nelson model~\cite{Hatano1998-jg,Hatano1996-sw}, a paradigmatic non-Hermitian model with asymmetric hopping. 
Although the effective non-Hermitian Hamiltonian associated with our Lindblad equation is equivalent to the Hatano--Nelson model, the spectrum and eigenmodes of the full Liouvillian are generally different from those of the effective non-Hermitian Hamiltonian because the full Lindblad dynamics also contains quantum-jump terms. It is also known that Liouvillian skin effects can occur in open quantum systems and are often manifested through the localization of nonzero Liouvillian eigenmodes and the resulting relaxation dynamics\cite{Song2019,Haga2021,Yang2022}. In the present work, however, we focus on the zero mode of the Liouvillian, namely the steady state, rather than on the relaxation modes. We find that the bulk observables in the steady state of our model are insensitive to the boundary condition~\cite{Supple}. Consequently, the periodic boundary condition analysis gives the same bulk results as the open boundary condition analysis.
Below we analyze the steady state of this model for $\gamma^{(l)} = \gamma^{(g)} = h$ and its effective temperature. 

Finally, we interpret $\tilde{\mathcal D}_{\rm eff}$ as an effective current source and introduce an additional thermal dissipator $\mathcal D_{\rm th}$. The latter is chosen such that the steady state at $h=0$ is the thermal equilibrium state. The resulting model is
\begin{align}
\m{D}_{\mr{th}}[\rho]& = \sum_{k, k'} \xi_{\epk\rightarrow\epkp}\Bigl[L^{(\mr{th})}_{kk'}\rho(t)L^{(\mr{th})\dagger}_{kk'} - \frac{1}{2}\{L^{(\mr{th})\dagger}_{kk'}L^{(\mr{th})}_{kk'}, \rho(t)\}\Bigr].\\
L^{(\mr{th})}_{kk'}& = c_{k'}^{\dagger} c_{k}\label{thermal_jump}
\end{align}
where $\xi_{\epk\rightarrow\epkp}$ is the transition rate from the state with energy $\epk$ to the state with energy $\epkp$ without the change of particle number caused by the thermal bath, and it satisfies the Kubo--Martin--Schwinger condition as $\xi_{\epk\rightarrow\epkp} = \xi_{\epkp\rightarrow\epk} \exp[-\beta(\epkp-\epk)]$ where $\beta$ is the inverse temperature of the thermal bath.
Our model is described by the following Lindblad equation as
\begin{align}
\dt \rho(t) =& -i[\m{H},\rho(t)] + \tilde{\m{D}}_{\mr{eff}}[\rho(t)] + \m{D}_{\mr{th}}[\rho(t)].\label{final_Lindblad}
\end{align}
In this setup, while $\m{D}_{\mr{th}}$ represents the thermal bath, $\tilde{\m{D}}_{\mr{eff}}$ represents the current source inherited from the boundary driving and $h$ is determined from the strength of the local current in the steady state of the original BDS. Thus, different from the analysis with electric field $E$, this model is directly described by the parameter $h$ related to the current density. 


\section{Steady-State Analysis}
In this work, we focus on the expectation value of current and the effective temperature, and it is sufficient to approximate the steady state as
\begin{align}
\rho_{\mr{ss}}&\simeq \prod_k\otimes\rho_{\mr{ss}}(k),\quad \rho_{\mr{ss}}(k) = p_k \ket{k}\bra{k} + (1-p_k)\ket{0} \bra{0}.\label{k_diag}
\end{align}
As a first step to analyze the steady state, we consider terms up to first order in $h$; the parameter should be determined from the strength of the local current in the steady state of the original model. We write the deviation from the Fermi distribution as
\begin{align}
    \delta p_k = p_k - f_k, \label{perturb}
\end{align}
where $f_k = 1/(1+\exp[\beta(\tilde{\epsilon_k})])$ is the Fermi distribution function, $\tilde{\epsilon_k} = \epk -\mu$, and $\beta$ is the inverse temperature of the thermal bath. Below we consider the small $h$ regime and analyze the steady state by perturbation theory with respect to $h$. Here, we summarize only the assumptions and main results of the calculation, while the detailed derivation is provided in the Supplemental Material\cite{Supple}.
By inserting Eq.~(\ref{perturb}) and Eq.~(\ref{k_diag}) into Eq.~(\ref{final_Lindblad}), we obtain the following equation for $\delta p_k$:
\begin{align}
    0=&-2h(1-\sk)f_k + 2h(1+\sk)(1-f_k)\nonumber\\
    & \  + \sum_{k'}\Bigl[\xi_{\epkp\rightarrow\epk}\bigl\{\delta p_{k'}(1-f_k) - f_{k'}\delta p_k \bigr\}\nonumber\\
    & \ \ \ \ \ \ \ \ -\xi_{\epk\rightarrow\epkp}\bigl\{\delta p_k(1-f_{k'}) - f_{k}\delta p_{k'} \bigr\}\Bigr],\label{ss_equation}
\end{align}
and then we get
\begin{align}
    \delta p_k &= \delta\rho^{(\mr{c})}_k + \delta\rho^{(\mr{th})}_k + \delta\rho^{(\mu)}_k,\label{results}\\
    \delta\rho^{(\mr{c})}_k&=h \tau_k \beta\sk\sech(\beta\tilde{\epk}/2)/\pi,\label{current_term}\\
    \delta\rho^{(\mr{th})}_k&=h \tau_k\beta\sech(\beta\tilde{\epk}/2)\tanh(\beta\tilde{\epk}/2)/\pi,\label{thermal_term}\\
    \delta\rho^{(\mu)}_k&=h \tau_k \beta\sech^2(\beta\tilde{\epk}/2)A_k/\pi,\label{chemical_potential_term}\\
    \frac{2\pi}{\beta\tau_k} &= \sum_{k'}\sqrt{\xi_{\epkp\rightarrow\epk}\xi_{\epk\rightarrow\epkp}}\sech(\beta\tilde{\epkp}/2),\label{relaxation_time}
\end{align}
where $A_k = \sum_{k'}\sqrt{\xi_{\epkp\rightarrow\epk}\xi_{\epk\rightarrow\epkp}}\delta\rho_{k'}\cosh(\beta\tilde{\epkp}/2)$.
We note that $\delta\rho_k^{(\mr{c})}$ is the current driving term of the steady state as $\ev{J} = \tr[J\rho_{\mr{ss}}] = \tr[J\rho^{(\mr{c})}_{\mr{ss}}]$, $\delta\rho_k^{(\mr{th})}$ represents heat emission into the bath in the steady state as $\ev{\dt Q_{s\rightarrow b}} = \tr[\dt Q_{s\rightarrow b}\rho_{\mr{ss}}] = \tr[\dt Q_{s\rightarrow b}\rho^{(\mr{th})}_{\mr{ss}}]$, and $\delta\rho_k^{(\mu)}$ represents the correction to chemical potential due to the coupling to the bath. 

In the zero-temperature limit, we use the approximation $\beta\sech(\beta\tilde{\epk}/2)\simeq 2\pi \delta(\tilde{\epk})$, and the expectation value of current reads as
\begin{align}
    \ev{J} &\simeq \sum_k \tau_k v_k^2\delta(\tilde{\epk}) \left(\frac{h}{t}\right)\simeq \tau D(\epsilon_F) v^2_F \left(\frac{h}{t}\right),\label{linear-like}
\end{align}
where $v_k = \p\epk/\p k$ is the velocity, $v_F = |\p\epk/\p k|_{\epk=0}$ is the Fermi velocity, $\tau = \tau_k|_{\epk=0}$, and $D(\epsilon_F)$ is the density of states at the Fermi level. 
Eq.~(\ref{linear-like}) reproduces the standard linear-response result upon identifying $h/t$ with the electric field $E$.
Moreover, approximating $\sinh(\beta\epk/2)\simeq \beta\epk/2$ and using $\beta\sech^2(\beta\epk/2)/4=-(\p f/\p\omega)_{\epk}$, we can derive
\begin{align}
    \delta\rho^{(\mr{th})}_k&\simeq \frac{2h\tau_k \tilde{\epk}}{\pi} \left(-\frac{\p f}{\p\omega}\right) \\
    &= -\frac{2\beta h\tau_k}{\pi} \left(\frac{\p f}{\p\beta}\right) = \delta\beta\left(\frac{\p f}{\p\beta}\right),\label{effT}
\end{align}
where $\delta\beta = -2\beta h \tau_k/\pi$, which reads $\delta T/T = 2 h\tau_k/\pi$. Thus, $\delta\rho_k^{(\mr{th})}$ can be considered as the increase of the effective temperature and it causes heat emission to the thermal bath. By using Eq.~(\ref{linear-like}) and Eq.~(\ref{effT}), we can derive the relation between the increase of the effective temperature and the current expectation value as
\begin{align}
    \frac{\delta T}{T} \simeq \frac{ W\ev{J}}{\pi v_F^2D(\epsilon_F)},\label{T-Jrelation}
\end{align}
where $W = 2t$ is the half bandwidth of the system.
Thus, we obtain the linear dependence of effective temperature increase on the current.

The linear relation in Eq.~(\ref{T-Jrelation}) does not rely on any property specific to one dimension. It follows from the weak-driving expansion, in which both the current density and the effective-temperature increase are linear in the current-source strength $h$, together with the low-temperature approximations for the Fermi distribution and the relaxation time. The same argument therefore applies to two- and three-dimensional systems under the corresponding assumptions, as shown explicitly in the Supplemental Material~\cite{Supple}.

We note that, although Eq.(\ref{linear-like}) corresponds to the results of the linear response theory at $h/t = E$, this identification does not imply that $h$ is always proportional to $E$ and, in general, $h$ is the control parameter determined by the current in the steady state.

Finally, at low temperature, we can approximate $\delta\rho_k^{(\mu)}$ as 
\begin{align}
    \delta\rho_k^{(\mu)} &\simeq h\tau_k\sech^2(\beta\tilde{\epk}/2)A_k/\pi\\
    & \simeq \frac{2h\tau A}{\pi} \left(-\frac{\p f}{\p\omega}\right) = \delta\mu\left(-\frac{\p f}{\p\omega}\right).\label{eff_mu}
\end{align}
Because we can approximate $\sech(\beta\epk/2)\simeq 2\pi \delta(\tilde{\epk})/\beta$ at low temperature, and $\rho^{(\mr{th})}_k$ has the linear dependence of $\beta$ as in Eq.~(\ref{effT}), $A_k$ can be approximated as $A_k \simeq A\beta$ where the factor $A$ is determined by the self-consistent equation. Thus, $\delta\rho_k^{(\mu)}$ can be considered as the correction to the chemical potential due to the coupling to the bath.


\section{Summary and Discussion}
In this paper, we have introduced a translationally invariant asymmetric-hopping model as an effective model for the bulk steady state of the BDS under current, and analyzed the effective temperature and the current density in its steady state. We found that the effective temperature increase is linearly proportional to the current density, similar to the experimental observation in Ref.\cite{Ootsuki2025-em}. 

First, we briefly analyzed the boundary-driven system and checked that the jump terms only appear at the edges and the local distribution has higher effective temperature than the edge bath. Then, we have shown that, by focusing on the local bulk state under current, a one-particle local effective Liouvillian which has the steady state with finite local current expectation value and uniform particle distribution must contain an asymmetric hopping term in the jump operator. 

We have constructed the translationally invariant effective model by summing up the local two-site effective Liouvillian. 
In this construction, the control parameter $h$ of the asymmetric hopping term is determined from the strength of the local current in the steady state of the original BDS, and thus we can get the direct relation between the current density and the effective temperature increase by analyzing the steady state of this model.
We have analyzed the steady state of this model in the simplest setup where the effective non-Hermitian Hamiltonian corresponds to Hatano--Nelson model, and found that the effective temperature increase is linearly proportional to the current density. 

Our results imply that Hatano--Nelson type asymmetric hopping models, which have been extensively studied in the context of non-Hermitian physics, especially in optics, can emerge as the effective description of current-carrying steady state in the solid-state system. 
Such models may therefore provide a useful framework for analyzing the current and effective temperature in nonequilibrium steady states, complementing previous connections between asymmetric hopping and driven systems\cite{Hatano1996-sw, Fukui1998-ey, Takasan2025-ql, Cao2023-mh}.

The analytical results presented here are obtained for noninteracting fermions. The construction of the effective bulk model itself is based on local steady-state properties and can, in principle, be extended beyond this setting, although its application to interacting systems requires an appropriate description of their low-energy degrees of freedom. In particular, an extension based on renormalized quasiparticles may be possible in a weakly interacting Fermi-liquid regime, whereas non-Fermi-liquid systems such as one-dimensional Luttinger liquids require a different treatment. Extending the present framework to multiorbital systems is another interesting direction, where the interplay between current-induced nonequilibrium distributions and internal orbital degrees of freedom may provide a route to exploring the role of quantum geometry in the effective temperature.

We hope that our work will be a step toward understanding the effective temperature in the steady state under current, and the intrinsic current-induced phenomena in condensed matter physics.

\

\section*{Acknowledgements}
Y. M sincerely appreciates Akito Daido, and Shuntaro Sumita for their fruitful comments at early stages of this work. Y. M. is supported by KAKENHI Grant Nos. JP24K16983 and JP26K17104. 
The author has used ChatGPT to check the English grammar and improve the readability of the manuscript.

\bibliography{AnalyzeHeatingProblem1,BoundaryDrivenSystem1,Current-Induced1,HatanoNelson_Current1,Heating_UnderDiscussion1,MagneticManipulationByCurrent1,NonEquilibriumEngineering1, ULE1}

\clearpage

\renewcommand{\thesection}{S\arabic{section}}
\renewcommand{\theequation}{S\arabic{equation}}
\renewcommand{\thefigure}{S\arabic{figure}}

\setcounter{equation}{0}
\setcounter{figure}{0}
\setcounter{section}{0}

\onecolumngrid
\begin{center}
{\large
{\bfseries Supplemental Materials for \\ ``Asymmetric-Hopping Effective Theory of Current-Carrying Bulk Steady States and Their Effective Temperature'' }}
\end{center}

\vspace{10pt}

\section{Rough Analysis of the Temperature Increase due to the Joule Heating}
We provide a rough analysis of the temperature increase due to the Joule heating.
To estimate the steady-state temperature, we add the thermal bath to the driven system, and denote the thermal conductance between the system and the bath by $\kappa$.

At the steady state, the heat-generation rate due to the Joule heating, and the heat-emission rate to the bath must be balanced as
\begin{align}
    \ev{\dt Q_{Joule}} &= \ev{\dt Q_{s\rightarrow b}},\\
    \ev{\dt Q_{Joule}} &= \rho\ev{J}^2, \quad \ev{\dt Q_{s\rightarrow b}} = \kappa (T_{\mr{eff}} - T_{\mr{bath}}),
\end{align}
where $\rho$ is the resistivity, $T_{\mr{eff}}$ is the effective temperature of the system at the steady state, and $T_{\mr{bath}}$ is the bath temperature. Thus, we can get
\begin{align}
    T_{\mr{eff}} = T_{\mr{bath}} + \frac{\rho\ev{J}^2}{\kappa}.\label{temp_increase}
\end{align}
Although  the resistivity $\rho$ can depend on the effective temperature, its dependence is expected to be $\propto T_{\mr{eff}}$ in a phonon-scattering dominant regime, $\propto T^{2}_{\mr{eff}}$ in a Fermi liquid, $\propto \exp[\Delta/T_{\mr{eff}}]$ in an insulator with an energy gap $\Delta$, or independent of $T$ at the impurity-scattering dominant regime. In none of these cases, does Eq.~(\ref{temp_increase}) yield the linear dependence of the effective temperature increase on the current, which is observed in the experiment\cite{Ootsuki2025-em}. 
Thus, a simple Joule-heating picture based on linear response theory in an isolated system does not explain the experimental observation.

\section{Bath Setup for the Boundary-Driven System in the Main Text}
We set the bath parameters for the boundary-driven system as follows. The left and right baths are set to be the thermal baths with the same temperature $T_{\mr{B}} = 0.05$ and chemical potential $\mu_{\mr{l}} = -\mu_{\mr{r}} = 2.4$, and the coupling strength between the system and the baths is set to be $\gamma$. The bath occupied spectral function is chosen as $J(\omega) = 1/(1+\exp[(\omega-\mu)/T]) \times 1/(1+\tau^2\omega^2)$ where $\tau$ is the relaxation time, which is set to be $\tau = 0.002$ in our calculation. We note that, while we have analyzed the boundary-driven system with this specific bath setup, our main results do not depend on this specific bath setup, as long as it produces a finite local current expectation value and boundary-localized dissipation in the steady state.

\section{Equivalence of Two Definitions of the Local Effective Temperature}
\label{sec:local_temperature}

In a nonequilibrium state, different operational definitions of the local effective temperature do not generally give the same result. 
Here, we compare the definition used in the main text, based on fitting the local occupation distribution to a Fermi distribution, with an alternative definition based on the local energy density.

For the two-site, single-orbital, noninteracting subsystem considered here, there are only two single-particle energy levels, with occupations $n_{+}$ and $n_{-}$. 
Once the local particle number $N_{\rm loc}=n_{+}+n_{-}$ is fixed, only one independent degree of freedom remains in the energy-diagonal part of the local distribution. 
Because only one independent degree of freedom remains, and it determines the effective temperature in both the Fermi distribution fitting and the local energy density matching, the two definitions yield the same result.
Consequently, the two definitions uniquely give the same effective temperature in the present case.

We emphasize that this equivalence is specific to the simple local subsystem considered here and does not generally hold for larger, multiorbital, or interacting nonequilibrium systems, where additional independent degrees of freedom remain even after fixing the local particle number and energy.

We note that the agreement of these temperature definitions also does not imply that the local state is thermal. 
In the present steady state, the finite current is carried by the off-diagonal coherence in the local energy basis, which is not determined by the occupation distribution or the local energy density. 
Thus, the local state remains nonequilibrium even though the two definitions yield the same effective temperature.

\section{Proof of the Finite Asymmetric Hopping Term in the Effective Local Liouvillian}
We use the three results, which are (i) the jump terms only appear at the edges, (ii) the site-resolved particle density is uniform in the bulk, and (iii) the local current is finite and constant in the bulk to prove that the effective local Liouvillian which has the steady state with finite local current expectation value must have the asymmetric hopping term in the jump operator. Owing to these three results the local two-site steady state in the bulk is written as
\begin{align}
\rho^{i, i+1}_{\mr{ss}} = \begin{pmatrix}
    \ket{0} & \ket{i} & \ket{i+1} & \ket{i, i+1}
\end{pmatrix}\begin{pmatrix}
    p_0 & 0 & 0 & 0\\
    0 & n & r + ic & 0\\
    0 & r - ic & n & 0\\
    0 & 0 & 0 & p_d
\end{pmatrix}\begin{pmatrix}
    \bra{0}\\ \bra{i} \\ \bra{i+1} \\ \bra{i, i+1}
\end{pmatrix},\label{local_steady_state}
\end{align}
where $\ket{0}$ is the vacuum state, $\ket{i}$ and $\ket{i+1}$ are the single-particle states at site $i$ and $i+1$, and $\ket{i, i+1}$ is the two-particle state at site $i$ and $i+1$. $p_0$, $n$, $r$, $c$, and $p_d$ are real numbers. The local current expectation value is given by $\ev{J} = 2tc$. Thus, if the local current expectation value is finite, $c$ must be finite. We then consider the local jump terms of the form
\begin{align}
    \m{D}_{i, i+1}[\rho] =& \sum_{p,q=i, i+1}\Biggl[ \gamma^{(l)}_{pq}\Bigl(c_{p} \rho c_q^{\dagger} - \frac{1}{2} \{c_q^{\dagger} c_p, \rho\}\Bigr) + \gamma^{(g)}_{pq}\Bigl(c_{p}^{\dagger} \rho c_q - \frac{1}{2} \{c_q c_p^{\dagger}, \rho\}\Bigr)\Biggr].
\end{align}
Because we consider the Lindblad form of the local jump term, the coefficients satisfy
\begin{align}
\gamma^{(l)}_{i,i+1} &= \gamma^{(l)*}_{i+1,i}, \quad \gamma^{(g)}_{i,i+1} = \gamma^{(g)*}_{i+1,i}, \quad \gamma^{(l)}_{i,i} = \gamma^{(l)*}_{i,i}, \quad \gamma^{(g)}_{i+1,i+1} = \gamma^{(g)*}_{i+1,i+1},\\
\gamma^{(l)}_{i,i}\gamma^{(l)}_{i+1,i+1} &\geq |\gamma^{(l)}_{i,i+1}|^2, \quad \gamma^{(g)}_{i,i}\gamma^{(g)}_{i+1,i+1} \geq |\gamma^{(g)}_{i,i+1}|^2.
\end{align}
When these jump terms and the local Hamiltonian have the local steady state in Eq.~(\ref{local_steady_state}), it satisfies
\begin{align}
    0 = \dt \rho^{i, i+1}_{\mr{ss}} &=-i[\m{H}^{i, i+1}, \rho^{i, i+1}_{\mr{ss}}] + \m{D}_{i, i+1}[\rho^{i, i+1}_{\mr{ss}}],\\
    \m{H}^{i, i+1} &= -t(c_i^{\dagger} c_{i+1} + c_{i+1}^{\dagger} c_i).
\end{align}
Focusing on the off-diagonal element $\ket{i}\bra{i+1}$, we can get
\begin{align}
    0 =& -\frac{1}{2}(\gamma^{(l)}_{i,i} + \gamma^{(l)}_{i+1, i+1} + \gamma^{(g)}_{i,i} + \gamma^{(g)}_{i+1, i+1})(r + ic) - (\gamma^{(l)}_{i,i+1} + \gamma^{(g)}_{i+1, i})n  + (\gamma^{(g)}_{i, i+1}p_0 + \gamma^{(l)}_{i+1, i}p_d),\label{off-diagonal}\\
    \Rightarrow & -\frac{1}{2}(\gamma^{(l)}_{i,i} + \gamma^{(l)}_{i+1, i+1} + \gamma^{(g)}_{i,i} + \gamma^{(g)}_{i+1, i+1})c - \mr{Im}[\gamma^{(l)}_{i,i+1} + \gamma^{(g)}_{i+1, i}]n  + (\mr{Im}[\gamma^{(g)}_{i, i+1}]p_0 + \mr{Im}[\gamma^{(l)}_{i+1, i}]p_d) = 0.\label{imaginary_part}
\end{align}
If $c$ is finite, $\mr{Im}[\gamma^{(l)}_{i,i+1} + \gamma^{(g)}_{i+1, i}]$ or $\mr{Im}[\gamma^{(g)}_{i, i+1}]p_0 + \mr{Im}[\gamma^{(l)}_{i+1, i}]p_d$ must be finite. Thus, the asymmetric hopping terms in the jump operator must be finite to have the local steady state with finite local current expectation value when site-resolved particle density is uniform.

\section{Validity of Choice of the Jump Operator in Section \ref{sec:model}}
Assuming that the coefficients $\alpha_i$ in the effective local Liouvillian operator have the same norm, the Liouvillian operator which cause the asymmetric hopping term can be written as
\begin{align}
L_j^{(l)} =&  a^{(l)}(c_j + e^{i\alpha}c_{j+1}), \quad L_j^{(g)} = a^{(g)}(c_j^{\dagger} + e^{-i\tilde{\alpha}}c_{j+1}^{\dagger}).\label{jump_op_current2}
\end{align}
When we consider the real part in Eq.~(\ref{off-diagonal}), we can get
\begin{align}
    0 =& -\frac{1}{2}(\gamma^{(l)}_{i,i} + \gamma^{(l)}_{i+1, i+1} + \gamma^{(g)}_{i,i} + \gamma^{(g)}_{i+1, i+1})r - \mr{Re}[\gamma^{(l)}_{i,i+1} + \gamma^{(g)}_{i+1, i}]n  + (\mr{Re}[\gamma^{(g)}_{i, i+1}]p_0 + \mr{Re}[\gamma^{(l)}_{i+1, i}]p_d).\label{real_part}
\end{align}
By inserting Eq.~(\ref{jump_op_current2}) into Eq.~(\ref{imaginary_part}) and Eq.~(\ref{real_part}), we can get
\begin{align}
    \frac{c}{r} &= \frac{-\mr{Im}[\gamma^{(l)}_{i,i+1} + \gamma^{(g)}_{i+1, i}]n  + (\mr{Im}[\gamma^{(g)}_{i, i+1}]p_0 + \mr{Im}[\gamma^{(l)}_{i+1, i}]p_d)}{-\mr{Re}[\gamma^{(l)}_{i,i+1} + \gamma^{(g)}_{i+1, i}]n  + (\mr{Re}[\gamma^{(g)}_{i, i+1}]p_0 + \mr{Re}[\gamma^{(l)}_{i+1, i}]p_d)}\nonumber\\
    &= \frac{a^{(l)}\sin(\alpha)(p_d-n) + a^{(g)}\sin(\tilde{\alpha})(p_0-n)}{a^{(l)}\cos(\alpha)(p_d-n) + a^{(g)}\cos(\tilde{\alpha})(p_0-n)}.\label{c_over_r}
\end{align}
Because, in our boundary-driven model, $r$ satisfies $r = \tanh(\beta t/2)$ and $\beta t \ll 1$, $c/r \gg 1$ for the finite current expectation value, and thus $\tan(\alpha)$ and $\tan(\tilde{\alpha})$ must be large. Therefore, we can approximate $\alpha \simeq \pi/2$ and $\tilde{\alpha} \simeq \pi/2$, which corresponds to the choice of the jump operator in the main text.

\section{Boundary-Condition Dependence of the Steady State of the Asymmetric-Hopping Model}
\begin{figure}[h]
\includegraphics[width=0.46\linewidth]{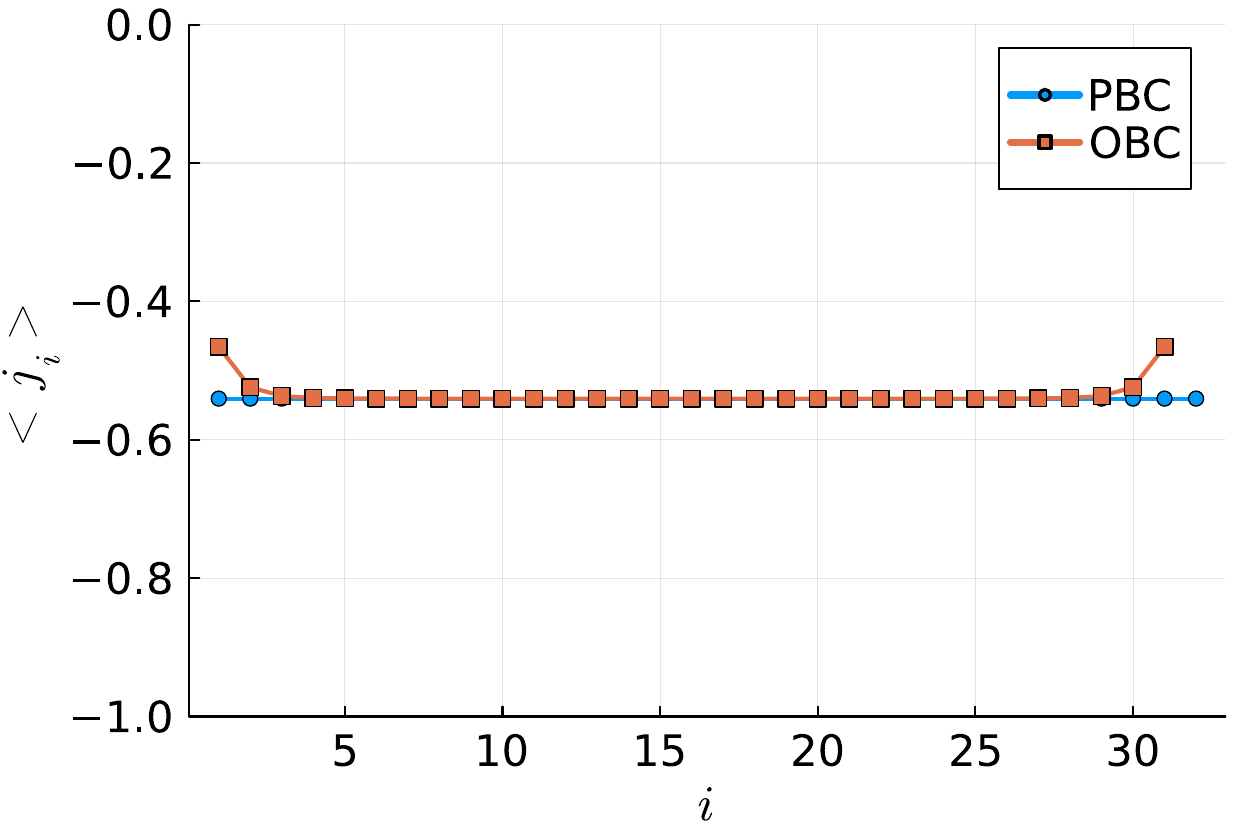}
\includegraphics[width=0.46\linewidth]{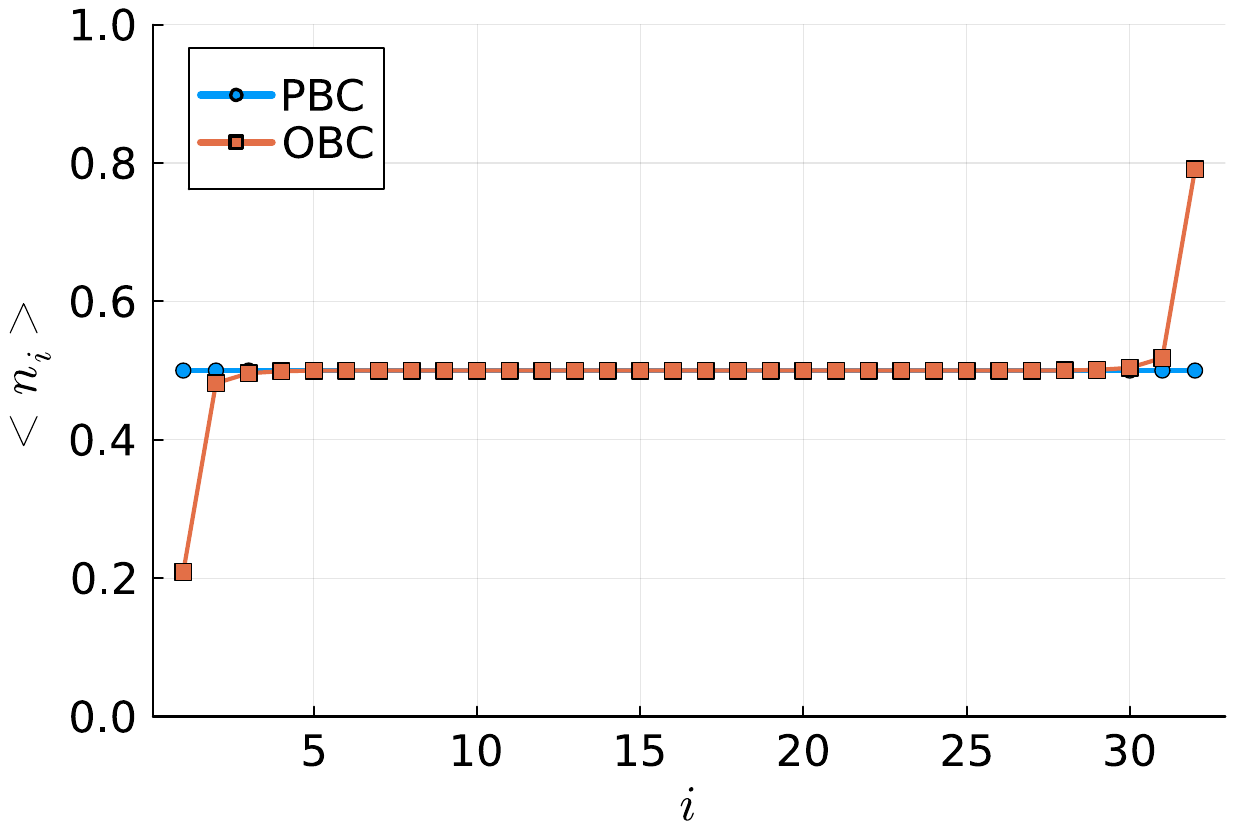}
\caption{The boundary-condition dependence of the steady state of the asymmetric-hopping model. The left and right panels show the current and site-resolved particle density at each site in the steady state under the periodic boundary condition and the open boundary condition, which shows that both expectation values are almost the same in the bulk region between these two boundary conditions. 
\label{fig:boundary_condition}}
\end{figure}

Here we numerically show that the bulk steady state of the asymmetric-hopping model is insensitive to the boundary condition. We analyze the steady state of the asymmetric-hopping model under the open boundary condition, and compare it with the steady state under the periodic boundary condition. Fig.~\ref{fig:boundary_condition} shows that the local distribution and the current expectation value in the bulk region are almost the same between these two boundary conditions, which implies that the bulk steady state of this model is insensitive to the boundary condition. This result supports our expectation that the effective asymmetric hopping model with periodic boundary conditions yields the same bulk steady state as the model with open boundary conditions, and thus can well describe the bulk steady state under current of the boundary-driven system with open boundary conditions.

\section{Derivation of the Steady State of the Asymmetric-Hopping Model}
Here, starting from Eq.~(\ref{ss_equation}), we will derive the steady state of the asymmetric-hopping model up to the first order of $h$. 
First, we divide $p_k^{(1)}$ into the symmetric part and the antisymmetric part as $p_k^{(1)} = \delta\rho_k^{(\mr{s})} + \delta\rho_k^{(\mr{a})}$, where $\delta\rho_k^{(\mr{s})} = (p_k^{(1)} + p_{-k}^{(1)})/2$ and $\delta\rho_k^{(\mr{a})} = (p_k^{(1)} - p_{-k}^{(1)})/2$.
Because $\epsilon_k = \epsilon_{-k}$ in our model, $f_k$, $\xi_{\epkp\rightarrow\epk}$, and $\xi_{\epk\rightarrow\epkp}$ are also symmetric, and thus Eq.~(\ref{ss_equation}) can be divided into the symmetric part and the antisymmetric part as
\begin{align}
    0 &= -2hf_k + 2h(1-f_k) + h\sum_{k'}\Bigl[\xi_{\epkp\rightarrow\epk}\bigl\{\delta\rho_{k'}^{(\mr{s})}(1-f_k) - f_{k'}\delta\rho_k^{(\mr{s})} \bigr\} -\xi_{\epk\rightarrow\epkp}\bigl\{\delta\rho_k^{(\mr{s})}(1-f_{k'}) - f_{k}\delta\rho_{k'}^{(\mr{s})} \bigr\}\Bigr]\\
    &= 2h\tanh\left(\frac{\beta\tilde{\epk}}{2}\right) + h\sum_{k'}\Bigl[\delta\rho_{k'}^{(\mr{s})}\left\{\xi_{\epkp\rightarrow\epk}(1-f_k) + \xi_{\epk\rightarrow\epkp}f_k\right\} - \delta\rho_k^{(\mr{s})}\left\{\xi_{\epkp\rightarrow\epk}f_{k'} + \xi_{\epk\rightarrow\epkp}(1-f_{k'})\right\}\Bigr]\\
    &= 2h\tanh\left(\frac{\beta\tilde{\epk}}{2}\right) + h\sum_{k'}\sqrt{\xi_{\epkp\rightarrow\epk}\xi_{\epk\rightarrow\epkp}}\Bigl[\delta\rho_{k'}^{(\mr{s})}\sech\left(\frac{\beta\tilde{\epk}}{2}\right)\cosh\left(\frac{\beta\tilde{\epkp}}{2}\right)- \delta\rho_k^{(\mr{s})}\sech\left(\frac{\beta\tilde{\epkp}}{2}\right)\cosh\left(\frac{\beta\tilde{\epk}}{2}\right)\Bigr],\\
    \Rightarrow & \delta\rho_k^{(\mr{s})} = \frac{\sech\left(\frac{\beta\tilde{\epk}}{2}\right)}{\sum_{k'}\sqrt{\xi_{\epkp\rightarrow\epk}\xi_{\epk\rightarrow\epkp}}\sech\left(\frac{\beta\tilde{\epkp}}{2}\right)}\Bigl[2 \tanh\left(\frac{\beta\tilde{\epk}}{2}\right) + A_k\sech\left(\frac{\beta\tilde{\epk}}{2}\right)\Bigr],\label{symmetric_ss}\\
    \nonumber\\
    0 &= 2h\sk + h\sum_{k'}\Bigl[\xi_{\epkp\rightarrow\epk}\bigl\{\delta\rho_{k'}^{(\mr{a})}(1-f_k) - f_{k'}\delta\rho_k^{(\mr{a})} \bigr\} -\xi_{\epk\rightarrow\epkp}\bigl\{\delta\rho_k^{(\mr{a})}(1-f_{k'}) - f_{k}\delta\rho_{k'}^{(\mr{a})} \bigr\}\Bigr]\\
    &= -2h\sk + h\delta\rho_k^{(\mr{a})}\cosh\left(\frac{\beta\tilde{\epkp}}{2}\right)\sum_{k'}\sqrt{\xi_{\epkp\rightarrow\epk}\xi_{\epk\rightarrow\epkp}}\sech\left(\frac{\beta\tilde{\epkp}}{2}\right)\\
    \Rightarrow & \delta\rho_k^{(\mr{a})} = \frac{\sech\left(\frac{\beta\tilde{\epk}}{2}\right)}{\sum_{k'}\sqrt{\xi_{\epkp\rightarrow\epk}\xi_{\epk\rightarrow\epkp}}\sech\left(\frac{\beta\tilde{\epkp}}{2}\right)}2\sk,\label{antisymmetric_ss}
\end{align}
where $A_k = \sum_{k'}\sqrt{\xi_{\epkp\rightarrow\epk}\xi_{\epk\rightarrow\epkp}}\delta\rho_{k'}^{(\mr{s})}\cosh(\beta\tilde{\epkp}/2)$.
Thus, we can get the results in Eq.~(\ref{results}) as
\begin{align}
    h p_k^{(1)} &= \delta\rho_k^{(\mr{a})} + \delta\rho_k^{(\mr{s})}\nonumber\\
    &= h\frac{\sech\left(\frac{\beta\tilde{\epk}}{2}\right)}{\sum_{k'}\sqrt{\xi_{\epkp\rightarrow\epk}\xi_{\epk\rightarrow\epkp}}\sech\left(\frac{\beta\tilde{\epkp}}{2}\right)}\Bigl[2\sk + 2 \tanh\left(\frac{\beta\tilde{\epk}}{2}\right) + A_k\sech\left(\frac{\beta\tilde{\epk}}{2}\right)\Bigr]
\end{align}
At low temperature limit, we can approximate $\beta\sech(\beta\tilde{\epk}/2)\simeq 2\pi \delta(\tilde{\epk})$, and $\sqrt{\xi_{\epkp\rightarrow\epk}\xi_{\epk\rightarrow\epkp}}\simeq \text{const}$ because the transition rates depend only on the energy $\epk$ and $\epkp$. We note that, when we approximate $\sinh(\beta\tilde{\epk}/2)\simeq \beta\tilde{\epk}/2$, $A_k$ must have linear dependence on $\beta$ because the first term of $\rho^{s}_k$ in Eq.~(\ref{symmetric_ss}) has the linear dependence of $\beta$, and thus we can approximate $A_k\simeq \beta A$ where $A$ is a constant.

\section{Generalization of the Current--Temperature Relation to Higher Dimensions}
\label{sec:higher_dimension}

Here we show that the linear relation between the effective-temperature increase and the current density obtained in the main text is not specific to one dimension. 
As the simplest examples, we consider a $d$-dimensional hypercubic lattice with the Hamiltonian
\begin{align}
    \m{H}&=-t\sum_{\bm{r}}\sum_{\alpha=1}^{d}\left(c_{\bm{r}}^\dagger c_{\bm{r}+\bm{e}_{\alpha}}+\mr{h.c.}\right),\quad \quad \quad \epsilon_{\bm{k}}
    =-2t\sum_{\alpha=1}^{d}\cos k_\alpha ,
\end{align}
where $\bm{e}_{\alpha}$ is the unit vector along the $\alpha$ direction.
The energy dispersion extends from $-2dt$ to $2dt$, and we define the half bandwidth as
$W=2dt$.
We assume that the current flows along the $x$ direction and extend the asymmetric-hopping dissipator in the main text along this direction. 
In momentum space, the corresponding dissipator is obtained from Eq.~(\ref{jump_current2}) by replacing $\sin k$ with $\sin k_x$.

In the weak-driving regime, the steady-state distribution can be expanded in the current-driving strength $h$ in the same manner as in the one-dimensional case.
The current-carrying and temperature-like corrections to the distribution are
\begin{align}
    \delta\rho^{(\mr{c})}_{\bm{k}}&=\frac{h\tau_{\bm{k}}\beta}{\pi}\sin k_x\sech\left(\frac{\beta\tilde{\epsilon}_{\bm{k}}}{2}\right),\label{current_term_d}\\
    \delta\rho^{(\mr{th})}_{\bm{k}}&=\frac{h\tau_{\bm{k}}\beta}{\pi}\sech\left(\frac{\beta\tilde{\epsilon}_{\bm{k}}}{2}\right)\tanh\left(\frac{\beta\tilde{\epsilon}_{\bm{k}}}{2}\right),\label{thermal_term_d}
\end{align}
where $\tilde{\epsilon}_{\bm{k}}=\epsilon_{\bm{k}}-\mu$.
No property specific to one dimension is used in obtaining Eqs.~(\ref{current_term_d}) and (\ref{thermal_term_d}).

First, using $\sinh(\beta\tilde{\epsilon}_{\bm{k}}/2)\simeq \beta\tilde{\epsilon}_{\bm{k}}/2$, we obtain, in the low-temperature regime,
\begin{align}
    \delta\rho^{(\mr{th})}_{\bm{k}}\simeq-\frac{2\beta h\tau_{\bm{k}}}{\pi}\frac{\partial f_{\bm{k}}}{\partial\beta}.
\end{align}
Assuming that $\tau_{\bm{k}}$ varies slowly within the thermally active energy window around the Fermi surface, we approximate it by its Fermi-surface value $\tau_F$. 
The temperature-like correction can then be expressed as
\begin{align}
    \delta\rho^{(\mr{th})}_{\bm{k}}\simeq\delta\beta\frac{\partial f_{\bm{k}}}{\partial\beta},\qquad\delta\beta=-\frac{2\beta h\tau_F}{\pi},
\end{align}
which gives
\begin{align}
    \frac{\delta T}{T}\simeq\frac{2h\tau_F}{\pi}.\label{deltaT_d}
\end{align}
Therefore, the temperature correction itself has no explicit dependence on the spatial dimension.

We next evaluate the current density along the $x$ direction. 
The group velocity is
\begin{align}
    v_x(\bm{k})=\frac{\partial\epsilon_{\bm{k}}}{\partial k_x}
    =2t\sin k_x.
\end{align}
Since only the antisymmetric correction $\delta\rho^{(\mr{c})}_{\bm{k}}$ contributes to the current, we obtain
\begin{align}
    \ev{J_x}&=\int_{\rm BZ}\frac{d^d\bm{k}}{(2\pi)^d}v_x(\bm{k})\delta\rho^{(\mr{c})}_{\bm{k}}
    =\frac{h\tau_F\beta}{2\pi t}\int_{\rm BZ}\frac{d^d\bm{k}}{(2\pi)^d}v_x^2(\bm{k})\sech\left(\frac{\beta\tilde{\epsilon}_{\bm{k}}}{2}\right).
\end{align}
Using the low-temperature approximation $\beta\sech(\beta\tilde{\epsilon}_{\bm{k}}/2)\simeq 2\pi\delta(\tilde{\epsilon}_{\bm{k}})$, we find
\begin{align}
    \ev{J_x}&\simeq\frac{h\tau_F}{t}\Phi_x(\epsilon_F),\label{J_d}\\
    \Phi_x(\epsilon_F)&\equiv\int_{\rm BZ}\frac{d^d\bm{k}}{(2\pi)^d}v_x^2(\bm{k})\delta(\epsilon_{\bm{k}}-\epsilon_F),\label{transport_DOS}
\end{align}
where $\Phi_x(\epsilon_F)$ is the transport density of states along the current direction.
Combining Eqs.~(\ref{deltaT_d}) and (\ref{J_d}), we obtain
\begin{align}
    \frac{\delta T}{T}\simeq\frac{2t}{\pi\Phi_x(\epsilon_F)}\ev{J_x}.\label{linear_relation_d}
\end{align}
Thus, the linear dependence of the effective-temperature increase on the current density remains valid in arbitrary spatial dimensions within the weak-driving and low-temperature regime considered here.

For an isotropic system, the transport density of states can be written as
\begin{align}
    \Phi_x(\epsilon_F)&=D(\epsilon_F)\ev{v_x^2}_{\rm FS}=\frac{D(\epsilon_F)}{d}\ev{\bm{v}^2}_{\rm FS},
\end{align}
where
\begin{align}
    \ev{\mathcal{O}}_{\rm FS}\equiv\frac{\displaystyle\int_{\rm BZ}\frac{d^d\bm{k}}{(2\pi)^d}\mathcal{O}(\bm{k})\delta(\epsilon_{\bm{k}}-\epsilon_F)}{\displaystyle D(\epsilon_F)}
\end{align}
denotes the Fermi-surface average.
Using $W=2dt$, Eq.~(\ref{linear_relation_d}) becomes
\begin{align}
    \frac{\delta T}{T}\simeq\frac{W}{\pi D(\epsilon_F)\ev{\bm{v}^2}_{\rm FS}}\ev{J_x}.\label{linear_relation_isotropic}
\end{align}
Therefore, the spatial dimension enters the proportionality coefficient only through the band structure and the Fermi-surface average of the velocity, while the linear dependence itself is independent of the spatial dimension.

Hence, no one-dimensional property is essential for the linear current-temperature relation derived in the main text. 
The detailed proportionality coefficient is determined by the transport density of states at the Fermi level and therefore generally depends on the band structure and the position of the Fermi level.

\end{document}